\documentclass[preprint]{aastex}
 
\newcommand{\be}{\begin{equation}} 
\newcommand{\ee}{\end{equation}}

\newcommand{\bea}{\begin{eqnarray}} 
\newcommand{\eea}{\end{eqnarray}} 
\newcommand{\eg}{e.g., } 
\newcommand{\ie}{i.e., } 
\newcommand{\OM}{\Omega_m} 
\newcommand{\OO}{\Omega_0}
\newcommand{\OL}{\Omega_{\Lambda}}

\begin{document}
\title{Distance-Redshift in Inhomogeneous $\OO=1$ 
Friedmann-Lema\^\i tre-Robertson-Walker Cosmology} 
\author{ R.  Kantowski }
\affil{ University of Oklahoma, Department of Physics and
Astronomy,\\ Norman, OK 73019, USA }
\email{kantowski@mail.nhn.ou.edu}

\author{ R. C. Thomas }
\affil{ University of Oklahoma, Department of Physics and
Astronomy,\\ Norman, OK 73019, USA }
\email{thomas@mail.nhn.ou.edu}

\begin{abstract} \vskip .2 truein 
Distance--redshift relations 
are given in terms of associated Legendre functions 
for partially filled beam observations in
spatially flat Friedmann-Lema\^\i tre-Robertson-Walker (FLRW) 
cosmologies. These models are dynamically pressure-free, flat FLRW on large scales
but, due to mass inhomogeneities, differ in their optical properties. The 
partially filled beam area-redshift equation is  a Lame$^{\prime}$
equation for arbitrary FLRW and is shown to simplify to the associated Legendre
equation for the spatially flat, \ie $\OO=1$ case.
We fit these new analytic Hubble curves to recent supernovae (SNe) data in an attempt to determine 
both the mass parameter $\OM$ and the beam filling parameter $\nu$. We find that current data 
are inadequate to limit $\nu$. However, we are able to estimate what limits 
are possible when the number of observed SNe is increased by factor of 10 or 100,
sample sizes achievable in the near future with the proposed SuperNova Acceleration 
Probe satellite.
\end{abstract}

\keywords{cosmology:  theory -- large-scale structure of universe}

\section{INTRODUCTION} \label{sec-intro} 
Distance-redshift or equivalently the Hubble curve is critical in determining
current values of the cosmological parameters $H_0, \OM$, and $\OL$. Conversely,
current values of these three parameters determine the large scale 
dynamics of the Universe into the distant past. A complication occurs when 
attempting to determine these parameters from high $z$ comparisons to the 
standard Hubble curve.  The standard Hubble curve is a theoretical quantity 
computed assuming all 
gravitating matter is homogeneously distributed; whereas, observational data 
is taken in the real inhomogeneous Universe. In an inhomogeneous universe
an observing light beam is lensed by inhomogeneities located external to,
but near the light beam, and defocused (relative to the standard Hubble curve) 
by the less than average matter density within the beam. 
The simplest way to take into account these effects is to
correct all beams for the missing homogeneous matter but correct 
for lensing only when necessary. This procedure requires the introduction of
one additional parameter, \eg a 
filling parameter 
$\nu, \ 0\le\nu\le2$ defined by the fraction of inhomogeneous 
matter $\rho_I/\rho_0 \equiv\nu(\nu+1)/6 
\le 1$ 
excluded from observing beams ($\nu=0$ is the standard 100\% filled beam FLRW
case and  $\nu=2$ is the empty beam case).
When observing high $z$ objects ($z\sim 1$) the reader can think of the parameter $\nu$  as
representing matter that exists in galaxies but not in the intergalactic medium. 
To find the theoretical Hubble curve for observations in such a universe one must solve the 
geometrical optics equation [see \cite{KR98}] given as equation (\ref{Area}) in the next section. 
This equation is actually equivalent to the Lame$^{\prime}$ equation 
for general FLRW but as pointed out by \cite{KKT} reduces to the 
associated Legendre equation (\ref{Legendre}) for the special 
case considered here, $\OO=1$.
In \S\,\ref{sec-lumdist} we solve this equation using appropriate boundary conditions
and give the Hubble curve in terms
of associated Legendre functions (eq.[\ref{Pans}]) as well as  
in terms of hypergeometric functions (eq.[\ref{2F1ans}]). 
In \S\,\ref{sec-fit} we fit this new Hubble curve to data for 60 supernovae (SNe)
from the Supernova Cosmological Project (SCP) and from the Cala$^{\prime}$n/Tololo
Supernova Survey (CTSS) in an attempt to determine 
the mass parameter $\OM$ and the filling parameter $\nu$. 
In \S\,\ref{sec-conclusions} we give some concluding remarks.

\section{The Luminosity Distance-redshift Relation} \label{sec-lumdist} 

For models being discussed here (and for most cosmological models), 
angular or apparent size distance 
is related to luminosity distance by $D_<(z)=D_{\ell}(z)/(1+z)^2$. 
We choose to give luminosity distances in this paper. 
The $D_{\ell}(z)$ which accounts
for a partially depleted mass density in the observing beam but neglects 
lensing by external masses
is found by integrating the second order differential equation for the 
cross sectional area $A(z)$
 of an observing beam from source ($z=z_s$) to observer ($z=0$), see \cite{KR98}
for some history of this equation,\footnote{This equation follows from 
applying Sach's optics equations [\cite{SR}] to an inhomogeneous FLRW
universe and neglecting  shear (external lensing) \cite{KR}. The first version
of equation (\ref{Area}) was 
given by \cite{Zel} and later \cite{DR74} included the cosmological term.}:
\bea 
&&(1+z)^3\sqrt{1+\OM z+\OL[(1+z)^{-2}-1]}\times\nonumber\\ &&\hskip 1 in {d\ \over
dz}(1+z)^3\sqrt{1+\OM z+ \OL[(1+z)^{-2}-1]}\,{d\ \over dz}\sqrt{A(z)}\nonumber\\ &&\hskip 2.0 in
+ {(3+\nu)(2-\nu)\over 4}\OM(1+z)^5\sqrt{A(z)}=0.   \label{Area} 
\eea
The required  boundary conditions are
\bea
\sqrt{A}|_s&=&0,\nonumber\\ {d\sqrt{A }\over dz}\Big|_s&=& -\sqrt{\delta\Omega} {c\over
H_s(1+z_s)}, \label{Aboundary} 
\eea 
 where $\delta\Omega$ is the solid angle of the beam at the source
and 
the FLRW value of the Hubble parameter at $z_s$ is
related to the current value $H_0$ at $z=0$ by 
\be
H_s=H_0(1+z_s)\ \sqrt{1+\OM
z_s+\OL[(1+z_s)^{-2}-1]}.  \label{Hs} 
\ee
The luminosity distance is then simply related to the area $A\big|_0$ of the beam 
at the observer by
\be
D_{\ell}^2\equiv {A\big|_0 \over \delta\Omega}(1+z_s)^2.
\label{Dl} 
\ee
Equation (\ref{Area}) can be put into the form of a Heun  equation
and its solution has been given in terms of Heun 
functions in \cite{KR98}.  Even though the Heun equation is only slightly more 
complicated than the hypergeometric
equation, \eg it has 4 regular singular points rather than 3, 
Heun functions are not yet available in standard libraries. 
Consequently, such expressions are not particularly 
useful for comparison with data, at this time. 
Because the exponents of three of the singular points of the 
area equation (in standard Heun form) are 0 and 1/2 
[see eq.\,(13) in \cite{KR98}], 
equation (\ref{Area}) is actually equivalent to the doubly periodic Lame$^{\prime}$
equation. We now show that it reduces to the associated Legendre equation
for the spatially flat universe, $\OO=1$. 
The required change of dependent and
independent variables
are respectively
\bea
P(A,z)&\equiv&(1+z)^{5/4}\sqrt{{A\over\delta\Omega}}\,,\\
\label{P}
\eta(z)&=&\sqrt{ {1+\OM z(3+3z+z^2)\over \OM(1+z)^3} }\,.
\label{eta}
\eea
The resulting associated Legendre equation is 
\be
(1-\eta^2){d^2P\over d\eta^2} -2\eta {dP\over d\eta} + 
\left(-\left[\frac{1}{6}\right]\left[\frac5{6}\right]-{[(1+2\nu)/6]^2
\over 1-\eta^2}\right)\ P = 0\ , 
\label{Legendre} 
\ee
with initial conditions: 
\bea
P|_s&=&0,\nonumber\\ 
{dP \over d\eta}\Big|_s&=&  {c\over
H_0}\frac23\ {\sqrt{\OM}\over 1-\OM}\ (1+z_s)^{11/4}. 
\label{Pboundary} 
\eea 
The resulting luminosity distance is then given by
\be
D_{\ell}(z_s)=(1+z_s)P(\eta(0)).
\label{DlP}
\ee
Expressed as associated Legendre functions equation (\ref{DlP}) becomes
\bea
&&D_{\ell}(\OM,\OL=1-\OM,\nu;z)= {c\over H_0}{
2\ \Gamma\left(\frac{5-2\nu}6\right)\Gamma\left(\frac{7+2\nu}6\right)(1+z)^{3/4}
\over 
(1+2\nu)\sqrt{\OM}
}\nonumber\\
&&
\times\Biggl[
{\rm P}^{\ (1+2\nu)/6}_{\ -1/6}\left(\sqrt{ {1+\OM z(3+3z+z^2)\over \OM(1+z)^3} }\right)
{\rm P}^{-(1+2\nu)/6}_{\ -1/6}\left({1\over \sqrt{\OM}}\right)
\nonumber\\
&&
-
{\rm P}^{\ (1+2\nu)/6}_{\ -1/6}\left({1\over \sqrt{\OM}}\right)
{\rm P}^{-(1+2\nu)/6}_{\ -1/6}\left(\sqrt{ {1+\OM z(3+3z+z^2)\over \OM(1+z)^3} }\right)
\Biggr].
\label{Pans}
\eea
In this expression the associated Legendre functions take on their
their analytically 
continued values (\ie the arguments are on the real axis and $> 1$).
When the filling parameter has values $\nu=0,1,$ or $2$, equation 
(\ref{Pans}) reduces respectively to
equations (22), (39) and (54), of \cite{KKT}. 
Because the associated Legendre equation is a special type of the hypergeometric equation 
it is always possible to write associated Legendre functions in terms of hypergeometric functions.
And because hypergeometric functions are the more universally available, these results 
are the more useful for most parameter values. That 
the hypergeometric result existed has independently been seen by \cite{KKT} and \cite{DM}.   
That the area equation  reduces to the
associated Legendre equation  for another special case ($\Lambda=0$) has been known 
for some time, see \cite{KVB} and \cite{SS}. 
The appropriate change of variables is
\bea
h(A,z)&\equiv&(1+z)\sqrt{{A\over\delta\Omega}}=(1+z)^{-1/4}P,\\
\label{h}
\zeta(z)&=& {\OM \over 1-\OM}(1+z)^3+1 = {\eta^2\over \eta^2-1}.
\label{zeta}
\eea
The resulting hypergeometric equation is 
\be
(1-\zeta)\zeta\ {d^2h\over d\zeta^2} +\left(\frac12-\frac76\zeta\right) {dh\over d\zeta} + 
{(\nu)(\nu+1)\over 36}\ h = 0\ , 
\label{hypergeometric} 
\ee
with initial conditions 
\bea
h_s&=&0,\nonumber\\ 
{dh \over d\zeta}\Big|_s&=&  -{c\over
H_s}\ {1-\OM\over 3\OM}\ (1+z_s)^{-2}.
\label{Fboundary} 
\eea 
The resulting luminosity distance is then given by
\be
D_{\ell}(z_s)=(1+z_s)h(\zeta(0)).
\label{DlF}
\ee
Expressed in terms
of hypergeometric functions equation (\ref{DlF}) becomes
\bea 
&&D_{\ell}(\OM,\OL=1-\OM,\nu;z)= 
{c\over H_0}{(1+z)\ 2\over (1+2\nu)\OM^{1/3}}\left[1+\OM z(3+3z+z^2)\right]^{\frac{\nu}6}
\nonumber\\
&&
\times
\Biggl\{
{}_2F_1\left( -\frac{\nu}6,\frac{3-\nu}6;\frac{5-2\nu}6;{1-\OM\over \left[1+\OM z(3+3z+z^2)\right]}\right) 
{}_2F_1\left( \frac{1+\nu}6,\frac{4+\nu}6;\frac{7+2\nu}6;1-\OM\right)
\nonumber\\
&&
-\ \left[1+\OM z(3+3z+z^2)\right]^{-\frac{1+2\nu}6}
{}_2F_1\left( -\frac{\nu}6,\frac{3-\nu}6;\frac{5-2\nu}6;1-\OM\right)\times
\nonumber\\
&&
{}_2F_1\left( \frac{1+\nu}6,\frac{4+\nu}6;\frac{7+2\nu}6;
{1-\OM\over\left[1+\OM z(3+3z+z^2)\right]}\right)
\Biggr\}.
\label{2F1ans}
\eea

\section{Prospects for constraining $\nu$ from high redshift SNe Ia} 
\label{sec-fit}
Recent cosmic microwave background observations strongly imply a
spatially flat universe (\cite{dB}, \cite{RH}), which naturally motivates
application of the 
new distance formulae presented in \S2 to a set of standard candles
to estimate $\nu$.  We use the 60 SNe Ia from the combined
Calan/Tololo + Supernova Cosmology Project (CT+SCP) as presented in
\cite{RA}, \cite{PS1}.  Combining these data with those from the
High-z SN search (\cite{SB}) would bring the total to about
100 SNe.  However, we shall see that this somewhat complicated task
would not increase the numbers of SNe enough to noticeably 
improve the estimate.

Rather than subject the data to an in-depth Bayesian re-analysis with
the additional beam filling parameter $\nu$ included, we merely use a
$\chi^2$ goodness-of-fit estimation.  We assume an intrinsic SN absolute
magnitude of $M_B=-19.33$ and $H_0=65$ km s$^{-1}$ Mpc$^{-1}$ and follow the
same procedure as \cite{WY2000a} to recover results consistent with
\cite{PS1} when $\nu=0$, i.e., for observations in a
homogeneous universe.  Then we employ the formulae presented here to
obtain confidence contours in the $\OM$--\,$\nu$ plane, constraining
$\OL$ so that $\OL=1.0-\OM$. By proceeding this way we are assuming that 
inhomogeneous matter, e.g., galaxies, are sufficiently removed from the lines of site 
of the 60 SNe and that lensing is negligible.

Figure 1 presents the 68, 90 and 99\% confidence contours for the fit
(solid, long-dashed and short-dashed lines respectively).  The results 
at $\nu=0$ are clearly consistent with the \cite{PS1}
findings.  Unfortunately, no value of $\nu$ can be ruled out with this
sample because of its size and depth in redshift space.  The best
fit overall is at ($\OM=0.31$, $\nu=0.0$). 

A simple-minded way to estimate the number of additional SNe required
to rule out any value of $\nu$ at 99\% confidence is to amplify the
contribution of each SN to $\chi^2$ by some factor.  This has the
effect of simulating more SNe which are exactly like the real ones,
and hence the actual best fit will not move, but the contours will
shrink.  Figures 2 and 3 are the same contours as in Figure 1, but
with sample size increased by factors of 10 and 100 respectively.  
We see that by simply enlarging the CT+SCP
sample 10 times will allow a $\nu=2$ value to be ruled out at 99\%
confidence.  A factor of 100 will allow a much larger range of $\nu$ to be
excluded.  These results are quite promising because a
sample size of SNe Ia in the thousands extending to even higher
redshifts should be possible with 
a ground based SNe pencil beam survey (\cite{WY2000b}) or a satellite mission such as
the SuperNova Acceleration Probe
(SNAP - http://snap.lbl.gov).

The fact that the current data at $z < 1$ do not rule out any value 
of $\nu$ is
not surprising.  Assuming ($\OM=0.3$, $\OL=0.7$), the
increase in distance modulus incurred by increasing $\nu$ from 0 to 2 at
$z=0.5$ is only 0.04 magnitudes (see \cite{KR98}).  However, at redshifts of $z=1.0$, 1.5 and
1.7 to be achieved in the future by SNAP, the increases are 0.14, 0.27
and 0.32 magnitudes respectively.  Lensing complications will begin to 
occur at these higher redshifts and will have to be corrected for and/or selected against
\{see \cite{WJ1}, \cite{PP1}, \cite{HD}, \cite{MS}, 
\cite{TPN}, and  \cite{WY1999}\}. 
Such data are likely to provide limited leverage in determining ($\OM$,
$\OL$), unless $\nu$ is properly constrained. Even though the most 
prevalent opinion is that $\nu=2$, i.e., there is no significant intergalactic medium, 
there may be a sea of massive neutrinos etc., out there that makes $0<\nu<2$. 

As observations reach $z\sim 3$ lensing by
galaxies becomes even more important and details of mass
inhomogeneities will significantly distort the Hubble curve. Distances given here 
are still useful even without exact knowledge of those inhomoheneities. 
A lower bound on the primary image magnification at a given redshift 
(relative to the mean) is given by 
$\mu_{cutoff} =D_{\ell}^2(\OM,\OL,\nu=0;z)/D_{\ell}^2(\OM,\OL,\nu;z).$ 
This bound represents sources which, by chance, are not lensed. 
This number can be compared to 
histograms given in  \cite{WJ2}, \cite{PP2}, \cite{HD}, and \cite{BGGM}.

\section{Conclusions} \label{sec-conclusions}

We have given useful forms for the luminosity distance in the currently
relevant inhomogeneous $\OO=1$ FLRW cosmologies.  These cosmologies are 
all dynamically FLRW in the large but differ in how gravitating 
matter affects optical observations.  A beam filling parameter 
$\nu, \ 0\le\nu\le2$ allows the matter to vary from completely 
transparent and homogeneous to completely inhomogeneous and exterior to 
any observing beam.
In order to determine the values of the cosmic parameters ($\OM$,
$\OL$) from Hubble curves at high redshift, the value of $\nu$
must also be constrained.  For fixed values of ($\OM,
\OL$), increasing $\nu$ from 0 (totally homogeneous universe) to
2 (totally clumped) increases the distance moduli of points on the
Hubble curve, {\it especially at higher redshifts} as pointed out in \S
\ref{sec-fit}.  When
observational error is taken into account, the problem of using
standard candles at high redshift while ignoring $\nu$ to obtain
$\OL$ will become particularly confounding.

The current sample of SNe Ia at $z < 1$ fail to constrain the value of 
the beam-filling parameter $\nu$.  Samples 10 to 100 times larger 
than the current sample, and in the same redshift
range, will constrain $\nu$.  In order to unambiguously determined 
($\OM$,$\OL$) from even higher redshift observations like 
those planned in the future, the distance-enhancing effect of $\nu$ must
be accounted for in the luminosity distance formulae.

\acknowledgements
The authors wish to thank Y. Wang, G. Kalbfleisch, David Branch, and E. Baron
for exceedingly helpful comments.

\clearpage

\figcaption[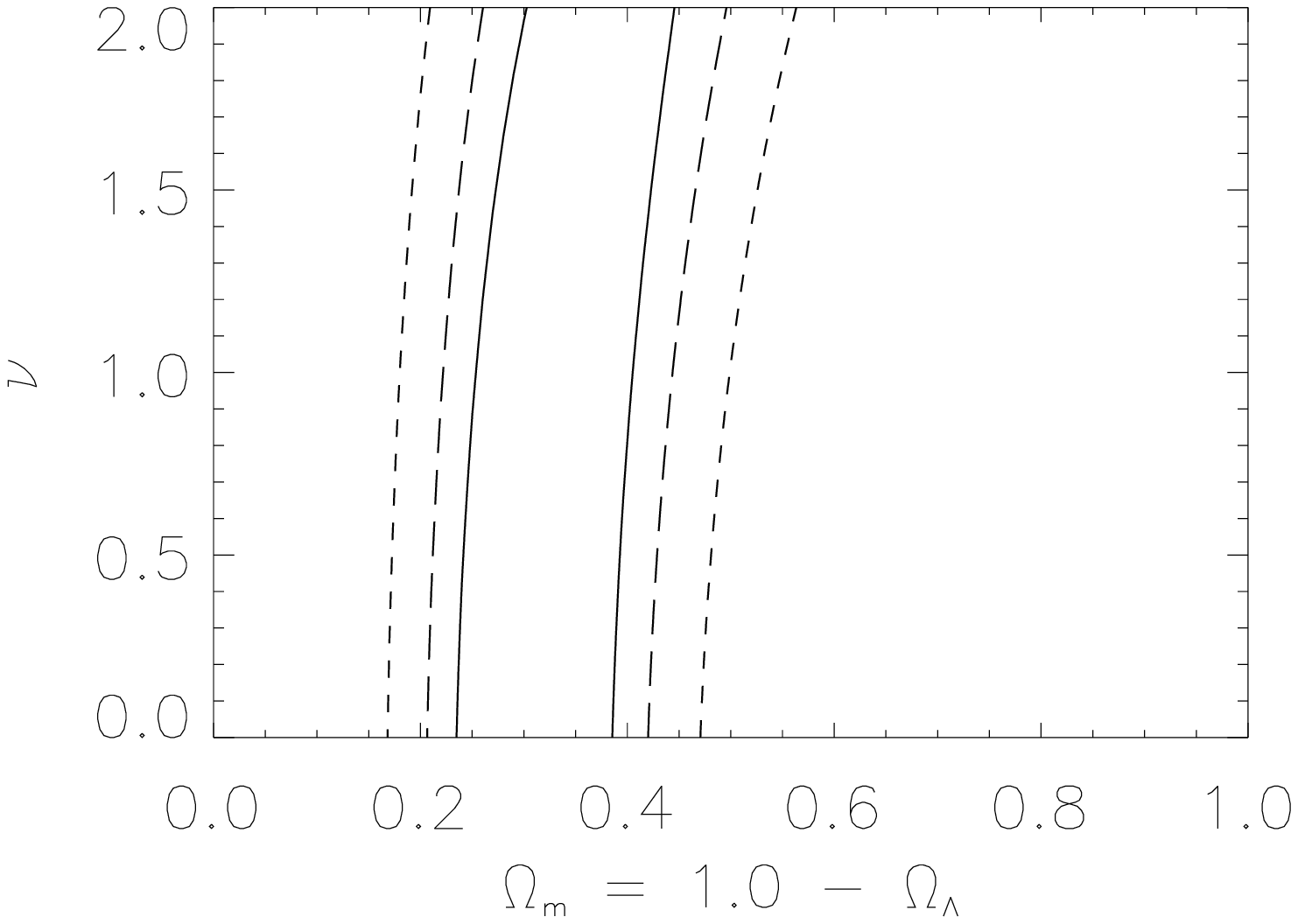]{The $\OM$--\,$\nu$ plane with 68, 90 and 99\%
confidence contours (solid, long-dashed and short-dashed, respectively) 
resulting from an attempt to constrain those parameters using the 60 SNe
Ia from the Calan/Tololo Supernova Search + Supernova Cosmology Project 
sample \cite{PS1}.  The data were first fit assuming $\nu=0$
(completely homogeneous universe) to recover a result consistent with
the original findings.  Then assuming $\OM = 1- \OL$ the above $\chi^2$
grid was calculated and contours of equal $\Delta \chi^2$ above the
minimum $\chi^2$ were plotted.
\label{fig1_3}}

\figcaption[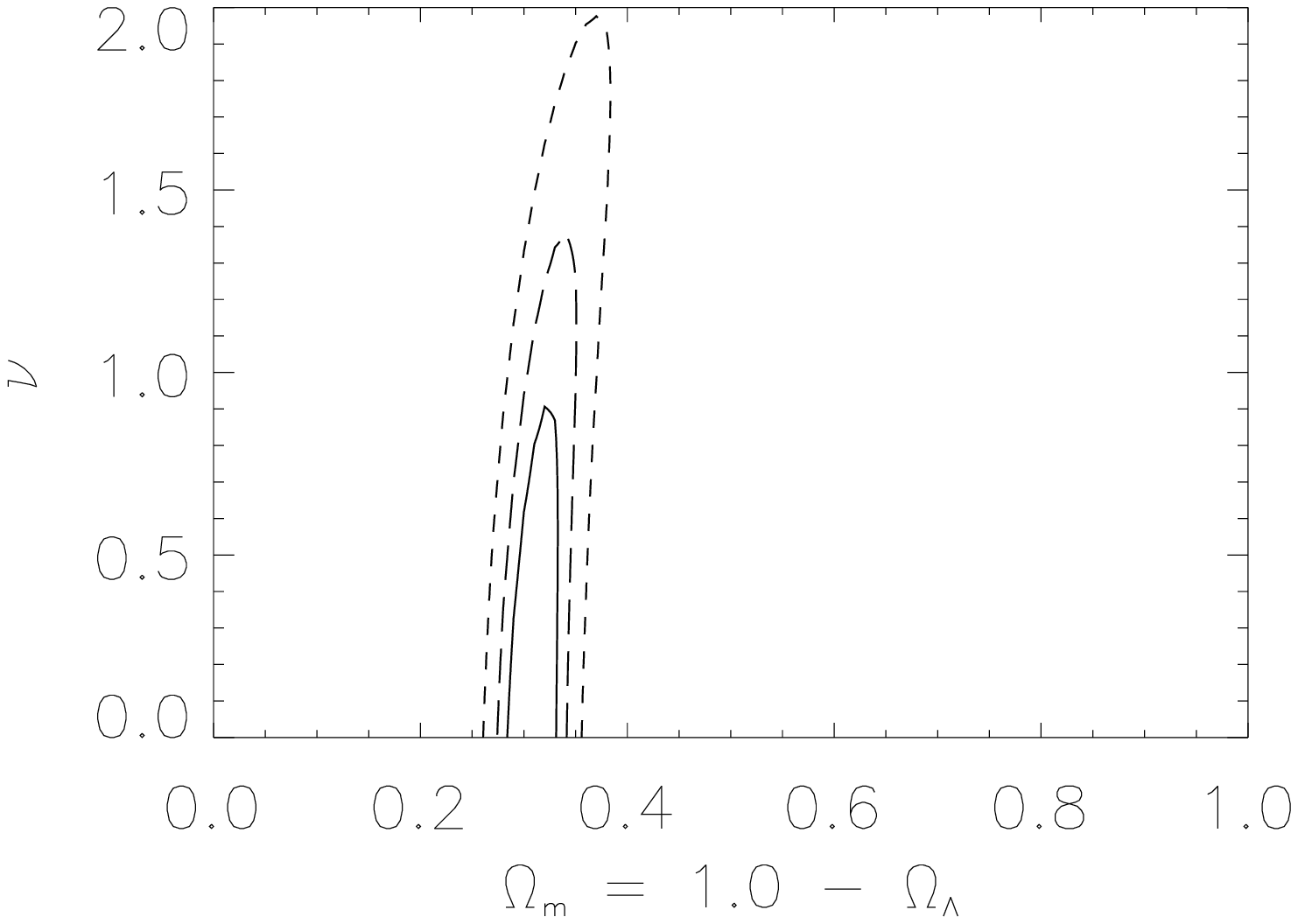]{Same as Fig.\,1 except 
10 times more SN measurements exactly like the 60 SNe Ia from the 
Calan/Tololo + Supernova Cosmology Project sample
were used in computing the confidence levels. 
\label{fig2_3}}

\figcaption[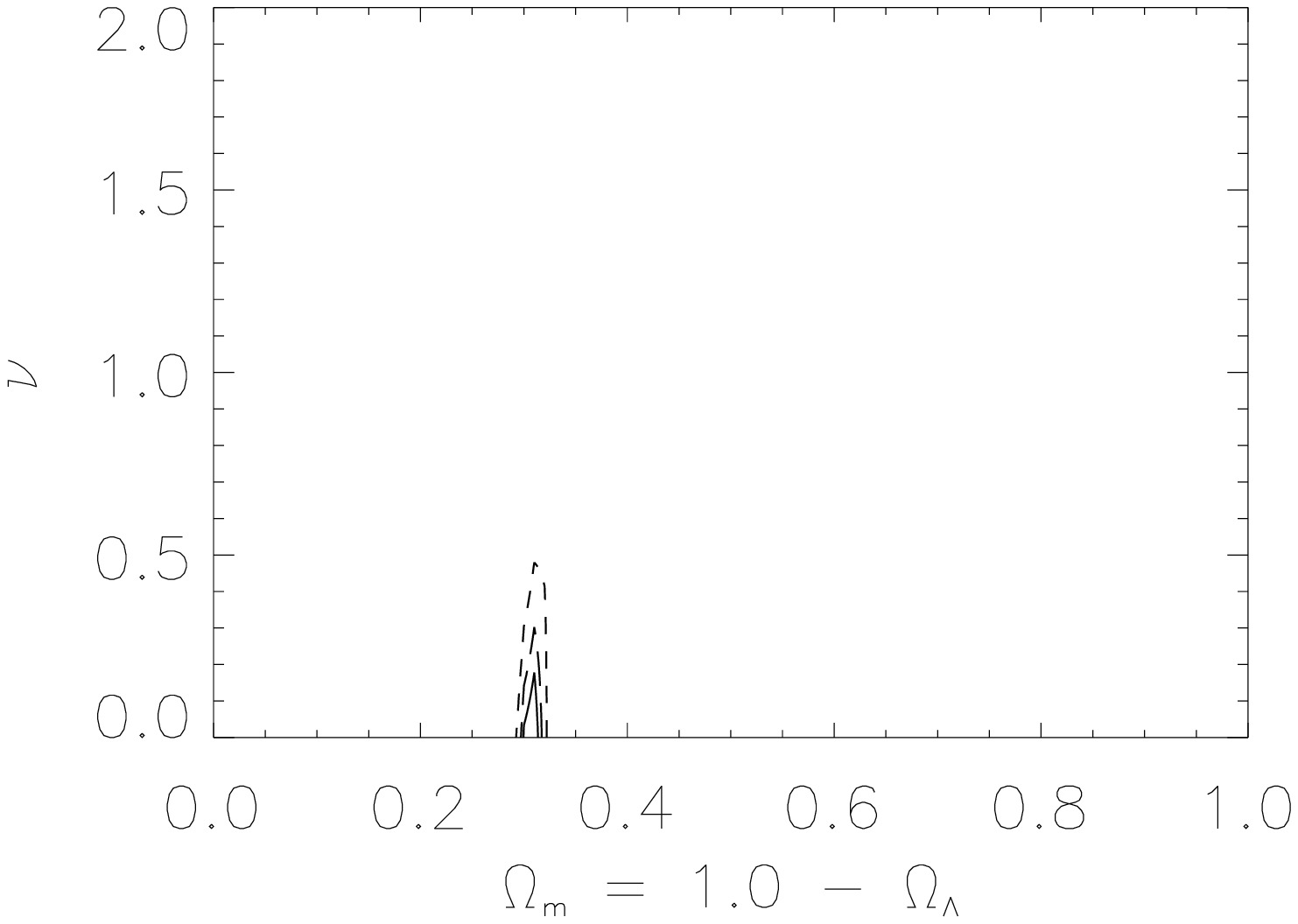]{Same as Fig.\,2 except 
100 rather than 10 times more SN were assumed to exist. 
\label{fig3_3}}

\plotone{fig1_3.eps}
\eject
\plotone{fig2_3.eps}
\eject
\plotone{fig3_3.eps}

\end{document}